\documentclass{elsart5}

\def\e{{\epsilon}}

\def\w{{\omega}}

 \usepackage{graphicx}

\usepackage{amssymb}

\begin{document}

\begin{frontmatter}

\title{Nonlocal Effect of Local Nonmagnetic Impurity 
in High-$T_{\rm c}$ Superconductors:\\
Induced Local Moment and Huge Residual Resistivity}

\author[aff1]{Hiroshi Kontani\corauthref{cor1}} 
\ead{kon@slab.phys.nagoya-u.ac.jp}
\corauth[cor1]{}
\author[aff1]{Masanori Ohno}
\address[aff1]{Department of Physics, Nagoya University,
Furo-cho, Nagoya 464-8602, Japan.} 
\received{May 2006}
\revised{2006}
\accepted{2006}


\begin{abstract}
We study a Hubbard model with a strong onsite impurity 
potential based on an improved fluctuation-exchange 
(FLEX) approximation, which we call the $GV^I$-FLEX method.
We find that (i) both local and staggered susceptibilities
are strongly enhanced around the impurity.
By this reason, (ii) the quasiparticle lifetime as well as the
local density of states (DOS) are strongly suppressed 
in a wide area around the impurity (like a Swiss cheese hole), 
which causes the ``huge residual resistivity'' beyond the s-wave 
unitary scattering value.
These results by the $GV^I$ method
naturally explains the various impurity effects in HTSC's
in a unified way,
which had been a long-standing theoretical problem.

\end{abstract}


\end{frontmatter}


In under-doped high-$T_{\rm c}$ superconductors (HTSC's),
nonmagnetic impurities (such as Zn) causes nontrivial 
widespread change of the electronic states.
For example, both the local and the staggered spin 
susceptibilities are strongly enhanced around the impurity,
within the radius of the AF correlation length $\xi_{\rm AF}$.
This behavior was experimentally observed in the 
site-selective $^{89}$Y NMR measurements for Zn-doped YBCO
 \cite{Alloul00-2}.
Moreover, a small concentration of Zn 
causes a huge residual resistivity, much beyond the 
s-wave unitary scattering limit
 \cite{Uchida}.
These nontrivial impurity effects had been
frequently considered as the evidence of the breakdown of the 
Fermi liquid state in under-doped HTSC's.

Pure HTSC's samples also show
various intrinsic non-Fermi liquid (NFL) behaviors
free from impurity effects.
Many of them had been explained based on the spin fluctuation theories
like the SCR theory and the fluctuation-exchange (FLEX) 
approximation
 \cite{Kontani-rev}.
For example, Curie like behavior of the Hall coefficient
$R_{\rm H}$ is naturally understood if one take the 
current vertex correction (CVC) due to AF fluctuations
into account \cite{Kontani-Hall}.
Anomalous transport phenomena in the pseudo-gap region,
such as a drastic increment of the Nernst coefficient,
are also reproduced by taking account of CVC
on the basis of the FLEX+T-matrix approximation
 \cite{Kontani-N}.

In the present paper,
we study a single impurity problem 
in a $(N\times N)$ square lattice Hubbard model (with $N=64$).
In general, it is not easy to obtain an appropriate solution
for this model because two different kinds of strong interactions 
have to be taken into account on the same footing.
To overcome this difficulty, we develop the $GV^I$-method, which is a 
powerful method of calculating the electronic states in real space.
Based on the $GV^I$-method,
we succeeds in explaining nontrivial impurity effects in HTSC's
{\it in a unified way}
within the scheme of the spin fluctuation theory,
without assuming any exotic mechanisms.


In the present model,
the self-energy $\Sigma({\bf r}-{\bf r}',\e)$
is not a function of ${\bf r}-{\bf r}'$
because of the absence of translational invariance.
Therefore, the self-energy in the site representation 
is expressed in a matrix form $N^2\times N^2$, ${\hat \Sigma}(\e)$.
In the present model, the self-energy
is divided into three parts;
${\hat \Sigma}^0(\e)$, impurity potential $I$, and 
$\delta{\hat \Sigma}(\e)$.
The first term corresponds to the self-energies
only by the Hubbard interaction $U$, and 
the last term is the cross-term between $U$ and $I$.

In the $GV^I$-method
 \cite{Kontani-imp},
${\hat \Sigma}^0$ is given by the FLEX approximation.
$\delta \Sigma_{i,j}(\e)$ is given by
$T\sum_{\w} G_{i,j}(\e+\w)V_{i,j}^I(\w)-{\hat \Sigma}^0$, 
where ${\hat G}=( \{{\hat G^{00}}\}^{-1}-{\hat \Sigma} )^{-1}$
and ${\hat G^{00}}(\e)$ is the noninteracting Green function.
Moreover, 
${\hat V}^I={\hat V}^{\rm FLEX}[{\hat G}^I ]$,
which is the effective interaction given by the FLEX-type
diagrams composed of 
${\hat G}^I=( \{{\hat G}^{00} \}^{-1}-{\hat I}-{\hat \Sigma}^0 )^{-1}$.
That is, ${\hat G}^I = {\hat G}|_{\delta\Sigma=0}$.
In the $GV^I$-method,
${\hat G}$ is solved self-consistently
whereas ${\hat V}^I$ is a partially self-consistent function.
As proved in ref. \cite{Kontani-imp},
$GV^I$-method is much {\it inferior} to a 
fully self-consistent $GV$-method.

\begin{figure}[t]
\includegraphics[scale =0.4]{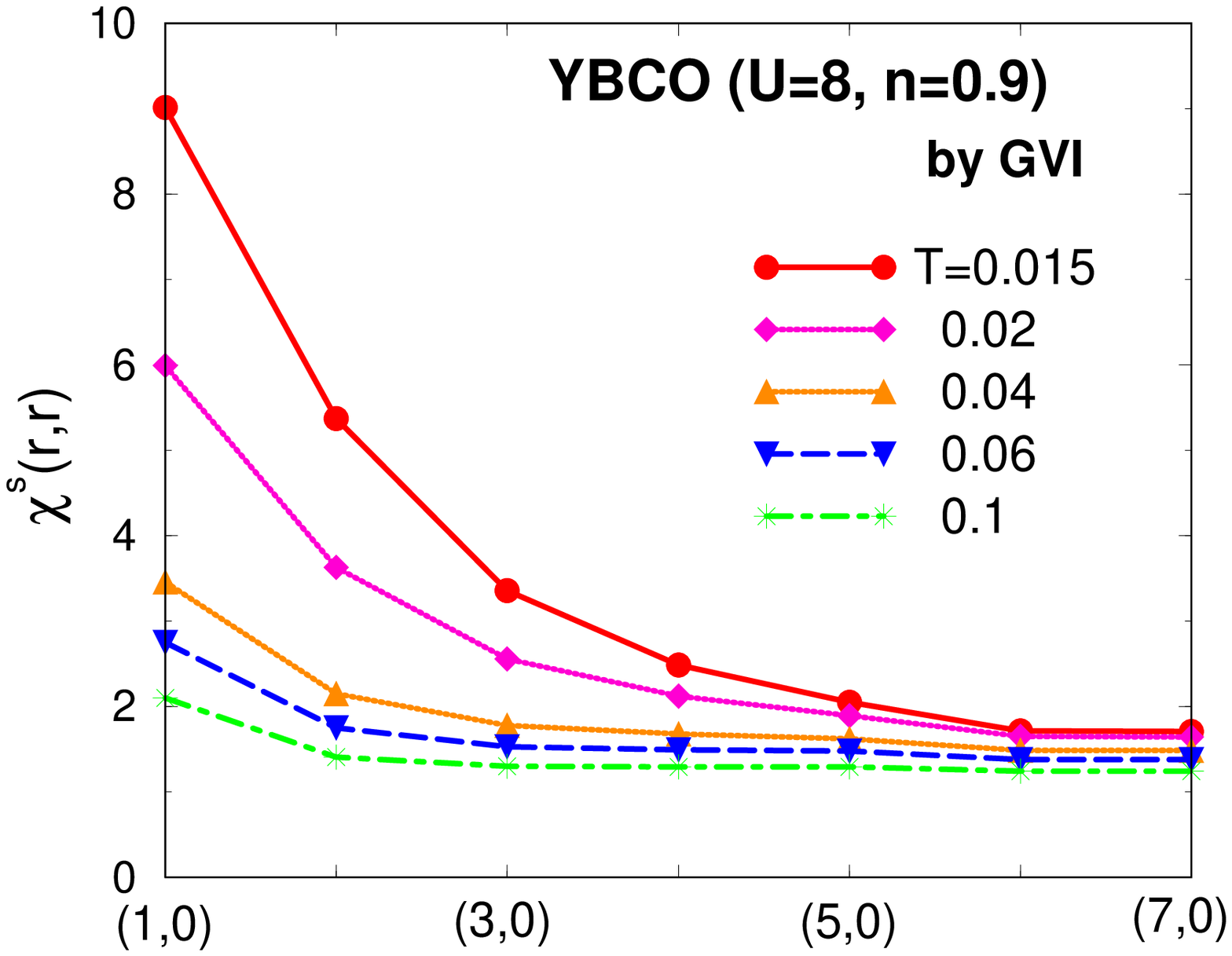}
\vspace{2mm}
\includegraphics[scale =0.408]{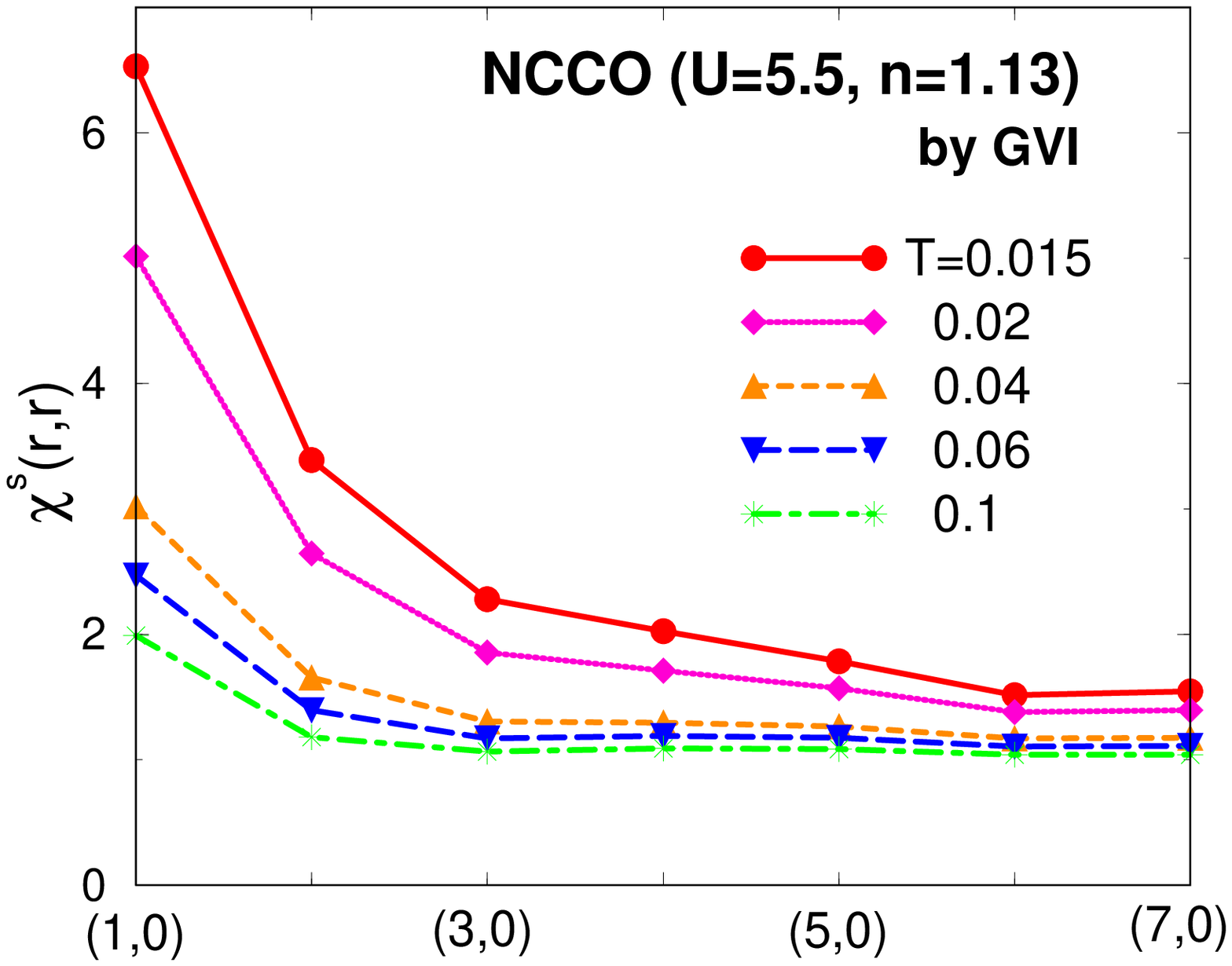}
\caption{Local spin susceptibility obtained by the $GV^I$-method
around the impurity site (at $(0,0)$).}
  	\label{fig1}
  \end{figure}

Hereafter, we take the unitary limit $I=\infty$.
Numerical results of the spin susceptibility 
given by the $GV^I$-method in site representation,
$\chi^{Is}({\bf r}, {\bf r})$, are explained in Ref.
 \cite{Kontani-imp}.
Here, we present the local susceptibility 
$\chi^{Is}({\bf r}, {\bf r})$ in Fig. \ref{fig1}, 
around the impurity site (at $(0,0)$) along the $x$-direction.
Here, we put $(t,t',t'')=(-1,1/6,-1/5)$, where 
$t'$, $t'$, and $t''$ are the nearest, the next nearest,
and the third nearest neighbor hopping integrals, respectively.
We assume $U=8$ for hole-doped system (YBCO)
and $U=5.5$ for electron-doped system (NCCO).
$T=0.02$ corresponds to 80K in real systems.
At ${\bf r}=(7,0)$, 
$\chi^{Is}({\bf r}, {\bf r})$ takes almost an bulk value without impurity.
However, it increases gradually as one approach the impurity.
The temperature dependence of $\chi^{Is}({\bf r}, {\bf r})$ 
away from the impurity is moderate.
On the other hand, it increases drastically around the impurity
as temperature decreases.
At the same time, the radius of the enlargement of 
$\chi^{Is}({\bf r}, {\bf r})$ increases.
The obtained result is highly consistent with NMR
measurements
 \cite{Alloul00-2}.

Here, we discuss the physical reason for the obtained impurity effect.
In the FLEX approximation, the AF-order (in the RPA) 
is suppressed by thermal and quantum fluctuations.
In the $GV^I$-method, the reduction of fluctuations due to an 
impurity gives rise to the enhancement of susceptibility.
However, this mechanism is absent in the RPA.
Therefore, the enhancement of susceptibility is tiny within the RPA.

\begin{figure}[t]
\includegraphics[scale =0.42]{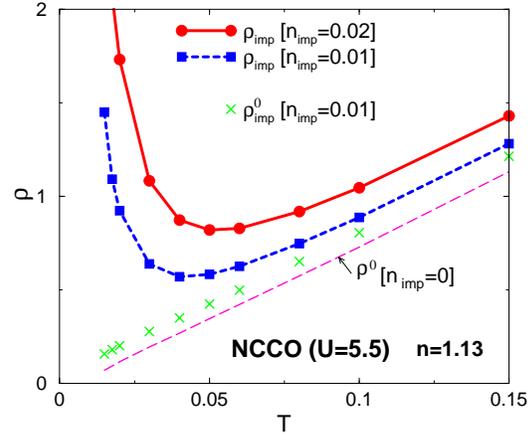}
\caption{Temperature dependence of $\rho$ for NCCO
with dilute impurities given by the $GV^I$-method.
}
  	\label{fig2}
  \end{figure}

Next, we discuss the transport phenomena in the presence 
of dilute impurities on the basis of the $GV^I$-method
 \cite{Kontani-imp}.
Figure \ref{fig2} shows the resistivity $\rho$ for NCCO
with $n_{\rm imp}=0$, 0.01 and 0.02.
The obtained result shows a huge parallel shift of resistivity
at finite temperatures ($\Delta\rho$) due to impurities,
far beyond the s-wave unitary scattering limit.
As $T$ decreases, nonmagnetic impurities cause a 
``Kondo-like upturn'' of $\rho$ below $T_x$,
reflecting an extremely short quasiparticle lifetime 
around the impurities.
We see that $T_x$ decreases as $n_{\rm imp}$ does.
The obtained Kondo-like upturn of $\rho$
strongly suggests that the insulating behavior of $\rho$ 
observed in under-doped HTSC's is caused by 
residual disorder in the CuO$_2$-plane, or 
residual apical oxygen in NCCO.

The present study reveals that a single impurity 
strongly affects the electronic states in a wide area around the impurity
in the vicinity of the AF-QCP.
Here, we developed the $GV^I$-FLEX method, which is a 
powerful method to study the impurity effect in strongly correlated systems.
Using the $GV^I$ method, characteristic impurity effects in under-doped 
HTSC's are well explained in a unified way, without introducing
any exotic mechanisms which assume the breakdown of the Fermi liquid state.
Qualitatively, these obtained numerical results are very similar
for YBCO, LSCO and NCCO.  
We succeeds in explaining nontrivial impurity effects in HTSC's
{\it in a unified way} in terms of a spin fluctuation theory,
which strongly suggests that the ground state of HTSC is a Fermi liquid.
We expect that novel impurity effects in other metals near AF-QCP,
such as heavy fermion systems and organic metals,
could be explained by the $GV^I$-method.


\end{document}